\documentclass[12pt]{article}
\usepackage{amsmath}
\usepackage{graphics,graphicx,epsfig}
\usepackage{amssymb}
\usepackage{epstopdf}
\usepackage{sidecap}
\usepackage{appendix}

\def\np{n^\prime}

\begin{document}

\begin{titlepage}
\vfill
\begin{flushright}
ACFI-T16-18
\end{flushright}

\vfill
\begin{center}
\baselineskip=16pt

{\Large\bf Building Cosmological Frozen Stars}
\vskip 0.15in
\vskip 10.mm

{\bf  David Kastor and Jennie Traschen} 

\vskip 0.4cm
Amherst Center for Fundamental Interactions, Department of Physics\\ University of Massachusetts, Amherst, MA 01003\\

\vskip 0.1 in Email: \texttt{kastor@physics.umass.edu, traschen@physics.umass.edu}
\vspace{6pt}
\end{center}
\vskip 0.2in
\par
\begin{center}
{\bf Abstract}
 \end{center}
\begin{quote}
Janis-Newman-Winicour (JNW) spacetimes generalize the Schwarzschild solution to include a massless scalar field.  Although suffering from naked singularities, they share the `frozen star' features of Schwarzschild black holes.
Cosmological versions of the JNW spacetimes were discovered some time ago by Husain, Martinez and Nunez and by Fonarev.  Unlike Schwarzschild-deSitter black holes, these solutions are dynamical, and the scarcity of  exact solutions for dynamical black holes in cosmological backgrounds motivates their further study.  Here we show how the cosmological JNW spacetimes can be built, starting from simpler, static, higher dimensional, vacuum `JNW brane' solutions via two different generalized dimensional reduction schemes that together cover the full range of JNW parameter space.  Cosmological versions of a BPS limit of charged dilaton black holes are also known.  JNW spacetimes represent a different limiting case  of the charged, dilaton black hole family.  We expect that understanding this second data point may be key to finding cosmological versions of general, non-BPS black holes.
\vfill
\vskip 2.mm
\end{quote}
\hfill
\end{titlepage}


\section{Introduction}

Questions about black holes in cosmology are of interest for a great many reasons.  For example, recent detections of gravity waves from the mergers of black hole binaries \cite{Abbott:2016blz} have led to the reconsideration of primordial black holes  \cite{zeldovich,Hawking:1971ei, Carr:1974nx} in the intermediate mass range as a dark matter candidate \cite{Bird:2016dcv,Clesse:2016vqa}.
Black holes in cosmology are dynamical, interacting with the matter that drives cosmic expansion.  We would like to understand to what extent such black holes can grow via the accretion of either cosmological fluids or fundamental matter fields, see {\it e.g.} the treatment in \cite{Jacobson:1999vr}.  More formal questions also arise,
such as how to formulate black hole thermodynamics in a such a non-equilibrium setting.

Exact solutions are important guides in general relativity.   However, there are few exact solutions known for black holes in cosmological backgrounds.  Analogues of the Kerr-Newman family of black holes in a deSitter background were found many years ago \cite{kottler,Carter:1968ks}.  However, these are stationary, at least within the deSitter horizon, and therefore non-dynamical\footnote{The deSitter multi-black hole solutions \cite{Kastor:1992nn,Brill:1993tm} are highly time-dependent.  However, they still do not display dynamical interactions of black holes with cosmological matter.}, reflecting  the fact that matter in the form of a cosmological constant cannot accrete.
Given the lack of actual dynamical, cosmological black hole solutions, it seems worthwhile to consider, as we do in this paper, a class of known non-stationary, inhomogeneous cosmological solutions, in which the inhomogeneities are not black holes, but do nonetheless share certain of their important physical features.
These solutions \cite{Husain:1994uj,Fonarev:1994xq} are cosmological versions of the static, asymptotically flat JNW family of solutions to Einstein gravity coupled to a massless scalar field \cite{Janis:1968zz}.

We will start by working with the action
\begin{equation}\label{action}
S=\int d^4 x\sqrt{-g}\left( R - {1\over 2}(\nabla\phi)^2 - V_0e^{-\lambda\phi}\right)
\end{equation}
for Einstein gravity coupled to a scalar field with an exponential potential.
The static JNW spacetimes \cite{Janis:1968zz} are solutions to this theory with $V_0=0$, {\it i.e.} with a massless scalar field, and are given by   
\begin{equation}\label{jnw}
ds^2 = -f^{\alpha}dt^2 +f^{-\alpha}dr^2 +f^{1-\alpha}r^2 d\Omega^2,\qquad 
e^\phi = f^{\pm\sqrt{1-\alpha^2}},\qquad  f = 1- {r_0\over r}
\end{equation}
Reality of $\phi$ constrains the parameter  $\alpha$ to the range\footnote{JNW spacetimes with parameters $\alpha$ and $\alpha^\prime =-\alpha$ are related by the coordinate transformation 
$r^\prime=r_0-r$ together with taking  $\phi^\prime=-\phi$.} $\alpha ^2\le 1$.
The JNW solution with  $\alpha=+1$  is  the Schwarzschild spacetime, which has a regular event horizon at $r=r_0$.
However, for $\alpha^2\neq  1$ the `would be event horizon' at $r=r_0$ is singular, as one would expect from the scalar no-hair theorems (see \cite{Herdeiro:2015waa,Sotiriou:2015pka} for recent reviews).
It follows from (\ref{jnw}) that the scalar field $\phi$ diverges at $r=r_0$,  as does its stress-energy as indicated by the scalar curvature, given by
\begin{equation}
R= {(1-\alpha^2)r_0^2\over 2 r^{2+\alpha} (r-r_0)^{2-\alpha}}
\end{equation}
Hence JNW solutions  with $\alpha^2 \neq 1$ are naked singularities rather than black holes. 
However, JNW spacetimes in the parameter range $0<\alpha<1$ are nonetheless similar to Schwarzschild in an important sense. One finds that the redshift factor for light traveling radially outward from an initial radius $r$ diverges as $r$ approaches $r_0$, and therefore like Schwarzschild these spacetimes can be regarded as ``frozen stars".

Physical properties of cosmological JNW spacetimes, such as the existence of black hole and cosmological apparent horizons,  have been studied in \cite{Husain:1994uj,Fonarev:1994xq} and further investigated in \cite{Maeda:2007bu}.  Our focus in this paper, however, is on the question of how cosmological solutions for black holes, or black-hole like objects in the JNW case, may be systematically obtained starting from non-cosmological solutions.  If we learn how to do this in the present case, we can then try to apply similar techniques to obtain actual dynamical, cosmological black hole solutions.
With this further application in mind, we will show that the cosmological JNW solutions \cite{Husain:1994uj,Fonarev:1994xq}  arise naturally,  via dimensional reduction from a simpler set of JNW-type brane solutions in $(4+n)$ dimensions.  We actually find two such constructions, one based on higher dimensional vacuum solutions and a second one based on higher dimensional solutions with a cosmological constant 
$\Lambda>0$.  We make use of ``generalized dimensional reduction"  \cite{Gouteraux:2011qh} in which the number 
$n$ of extra dimensions is allowed to vary continuously.
In each construction the cosmological JNW solution with parameter $\alpha$ descends from a particular number $n[\alpha]$ of (generalized) extra dimensions.  Together, the two constructions cover the full range of $\alpha^2\le 1$.
The fact that we have two constructions, based respectively on higher dimensional deSitter and vacuum solutions, is a cosmological analogue of the AdS/Ricci-flat correspondence  discussed in \cite{Caldarelli:2013aaa}.

The paper is organized as follows.  We begin in Section (\ref{hmnkk}) by presenting a particularly simple sub-case of our construction that relies only on familiar formulas and manipulations from $5D$ Kaluza-Klein theory.  Section (\ref{generalsection}) works through the general construction of cosmological JNW spacetimes.  We recall the generalized dimensional reduction strategy \cite{Gouteraux:2011qh} and show how it can be used in two ways to construct $4D$ FRW solutions for Einstein gravity coupled to a scalar field with an exponential potential.  The two methods cover complementary portions of the parameter space for such solutions.
We then present static, higher dimensional, vacuum `JNW brane' spacetimes, which serve as the starting point for the higher dimensional construction.   Finally, we present the two complementary generalized dimensional reduction schemes that produce the dynamical, inhomogeneous cosmological JNW spacetimes from static JNW-brane inputs and together cover the full range of JNW parameter space.   In Section (\ref{discussion}) we provide some discussion of our results and directions for future research.

\section{Warm-up in $D=5$}\label{hmnkk}

The cosmological generalization of the JNW spacetimes (\ref{jnw}) was found in two steps.  Husain, Martinez and Nunez (HMN)  \cite{Husain:1994uj} first found a cosmological version of (\ref{jnw}) restricted to the particular parameter values $\alpha = \pm\sqrt{3}/2$.  A few months later, Fonarev \cite{Fonarev:1994xq} found a more general cosmological solution that covers all values of $\alpha$ and includes the HMN spacetimes as a special case.  In both \cite{Husain:1994uj,Fonarev:1994xq}, the solutions are simply written down, apparently found by trial, and possibly error, using an inspired ansatz.  Our goal is to understand how these solutions can be built starting from the static JNW spacetimes.  In this section, we warm up for this task by showing how the dynamical HMN spacetimes may be obtained from the static  JNW spacetimes with $\alpha = \pm\sqrt{3}/2$ via simple manipulations in $D=5$ Kaluza-Klein theory.

The HMN spacetimes are again solutions to Einstein gravity coupled to a massless scalar field, {\it i.e.} to the theory described by the action (\ref{action}) with $V_0=0$, and are given by
\begin{equation}\label{hmn}
ds^2 = (t/t_0)\left[ -f^\alpha dt^2 + f^{-\alpha}dr^2 +r^2 f^{1-\alpha }d\Omega^2\right] ,\qquad
e^\phi =(t/t_0)^{\sqrt{3}}f^{{\alpha\over\sqrt{3}}}, \quad f= 1-{r_0\over r}
\end{equation}
where $\alpha = \pm\sqrt{3}/2$.
If we set the parameter $r_0=0$, this reduces to an FRW cosmology, with evolution is sourced by the time dependent massless scalar field\footnote{See the Appendix for a brief review of FRW-type solutions to the more general theory of Einstein gravity coupled to a scalar field with an exponential potental (\ref{action}).}.  With $r_0\neq 0$, these spacetimes are inhomogenious, but are clearly asymptotic in the regime $r\gg r_0$ to the massless scalar cosmology.  Moreover, if we consider (\ref{hmn}) restricted to a slice of constant time, or over a short period of time,  then we easily recognize the structure of the JNW spacetimes with $\alpha = \pm\sqrt{3}/2$.  It then seems reasonable to regard the HMN spacetimes (\ref{hmn}) as embeddings of JNW-type inhomogeneities with these values of $\alpha$ in the massless scalar FRW cosmology.

It is natural to wonder what picks out the particular values of $\alpha$ in the HMN solutions.  We will see that viewing the spacetimes (\ref{hmn}) as the dimensional reductions of $5D$ vacuum solutions provides a clear answer.
It is well known that a solution
 to Einstein gravity coupled to a massless scalar field  in $4D$ can be lifted, or oxidized, to  a solution of pure $5D$ Einstein gravity  by taking the $5D$ metric to be 
\begin{equation}\label{kk_diagonal}
ds_5^2 = e^{+2\phi/\sqrt{3}} dw^2 + e^{-\phi/\sqrt{3}}g_{\mu\nu}dx^\mu dx^\nu
\end{equation}
where $g_{\mu\nu}$ and $\phi$ are respectively the $4D$ metric and scalar field.  The $5D$ metric obtained in this way is necessarily invariant under translations in the new $w$-direction\footnote{The $4D$ U(1) gauge field that can also be present in this construction, will not play a role here.}.  Conversely, a $5D$ vacuum metric of this form, with $\phi$ and $g_{\mu\nu}$ independent of $w$, can be dimensionally reduced to a $4D$ solution of the Einstein-massless scalar system.   The HMN spacetimes, we will see, can be obtained through a series of steps, first lifting the
non-cosmological JNW spacetimes (\ref{jnw}) to $5D$, then making a certain $5D$ coordinate transformation, and finally reducing to $4D$ along a different translation invariant spatial direction.  We will see that the second step in the procedure works for the values of the JNW parameter $\alpha=\pm\sqrt{3}/2$ that appear in the HMN solutions. 

Before doing this, however, let us review how the FRW massless scalar cosmology, {\it i.e.} the HMN spacetime (\ref{hmn}) with $r_0=0$, can be obtained via dimensional reduction.  In this case, the process starts with  $5D$ Minkowski spacetime 
\begin{equation}\label{flat5d}
ds_5^2 = dy^2 -d\tau^2 + \delta_{ij}dx^idx^j,\qquad i,j=1,2,3
\end{equation}
Dimensional reduction along the $y$-direction using (\ref{kk_diagonal}) would simply yield $4D$ Minkowski spacetime and a trivial profile for the  scalar field.
However, an alternate reduction gives a 4D cosmology. We first transform the $y, \tau$ portion of the metric (\ref{flat5d}) to Milne coordinates by setting
\begin{equation}\label{milne1}
y=t\sinh (w/t_0),\qquad \tau=t\cosh (w/t_0)
\end{equation}
The flat $5D$ metric now has the form
\begin{equation}
ds_5^2 = ({t/t_0})^2 dw^2 - dt^2 +\delta_{ij}dx^idx^j
\end{equation}
Dimensionally reducing along the $w$-direction then yields the time dependent spacetime
\begin{equation}
ds_4^2 = (t/t_0)(-dt^2 + \delta_{ij}dx^idx^j),\qquad e^\phi = (t/t_0)^{\sqrt{3}}
\end{equation}
which we recognize as the massless scalar cosmology that appears as the $r_0=0$ limit of the HMN spacetimes (\ref{hmn}). Note that a similar transformation to Milne coordinates can be made whenever the $5D$ metric is independent of the coordinates $y$,  $\tau$ and contains the boost invariant $2D$ Minkowski factor $dy^2-d\tau^2$, possibly multiplied by a function of the other coordinates.  
It is this latter, boost symmetry of the $5D$ spacetime that picks out the particular values of the JNW parameter $\alpha$ in the HMN solutions.

In order to now build the HMN spacetimes, we start by lifting the JNW spacetimes (\ref{jnw}) to $5D$ using the Kaluza-Klein map (\ref{kk_diagonal}).   After relabeling the internal and time coordinates to be $y$, $\tau$ respectively, this yields the metric 
\begin{equation}\label{JNW_uplift}
ds_5^2 = f^{\pm2\sqrt{1-\alpha^2}/\sqrt{3}} dy^2 +f^{\mp\sqrt{1-\alpha^2}/\sqrt{3}}  \left( -f^{\alpha}d\tau^2 +f^{-\alpha}dr^2 +r^2 f^{1-\alpha}d\Omega^2\right)
\end{equation}
which is a solution to the vacuum Einstein equations and has translation symmetries in the $y$ and $\tau$ directions.
The two solutions with plus, or minus signs, come from
the $\pm$ options for the scalar field in the 4D JNW solution.  For $\alpha=1$, the $5D$ spacetime (\ref{JNW_uplift}) is simply the uniform black string.  For general $\alpha$ we simply call it a `JNW string'.   Similarly defined `JNW branes' with translation invariance in additional spatial worldvolume directions will be important in Section (\ref{generalsection}) below. 

In order to obtain a $4D$ cosmological solution via Kaluza-Klein reduction, we need to transform the $y$, $\tau$ coordinates in the $5D$ metric (\ref{JNW_uplift}) to Milne type coordinates (\ref{milne1}). 
Given the translation symmetries of the metric (\ref{JNW_uplift}), this will be possible if in addition 
the metric has the required boost symmetry.
The $5D$ metric (\ref{JNW_uplift}) will include the appropriate $2D$ Minkowski factor if $g_{\tau\tau}=-g_{yy}$, which will be the case if the JNW parameter $\alpha$  satisfies  $\alpha=\pm\sqrt{3(1-\alpha^2)}$, which is solved by
\begin{equation}\label{hmn-param}
\alpha=\pm{\sqrt{3}\over 2}
\end{equation}
These are precisely the values of $\alpha$ that appear in the cosmological HMN solutions (\ref{hmn}). The HMN solutions are indeed recovered, if after fixing $\alpha$ to one of these values one carries out the remaining steps in the procedure; transforming to Milne coordinates $w$, $t$ using (\ref{milne1}) and dimensionally reducing to $4D$ using the Kaluza-Klein ansatz (\ref{kk_diagonal}) to extract the $4D$ metric and massless scalar field.

\section{Building general cosmological JNW spacetimes}\label{generalsection}

In the last section, we saw how a $5D$ construction yields the HMN spacetimes, which embed JNW inhomogeneities with $\alpha=\pm\sqrt{3}/2$ in a massless scalar FRW cosmology.  In this section, we will use the generalized dimensional reduction scheme of reference \cite{Gouteraux:2011qh} to build up the general cosmological JNW spacetimes found by Fonarev \cite{Fonarev:1994xq}.
The Fonarev solutions \cite{Fonarev:1994xq} to the Einstein-scalar field system (\ref{action}) are given by
\begin{equation}\label{fonarev}
ds^2 = (t/t_0)^{2s}\left( -f^\alpha dt^2 +f^{-\alpha}dr^2 +f^{1-\alpha}r^2d\Omega^2\right),\qquad 
e^\phi= (t/t_0)^{2\epsilon\sqrt{s(s+1)}}f^{\pm\sqrt{1-\alpha^2}}
\end{equation}
where again $f(r)=1-r_0/r$ and the parameters of the scalar potential and the JNW parameter are given in terms of cosmological scale factor exponent by
\begin{equation}
\lambda=\sqrt{{s+1\over s}},\qquad V_0= {2s(2s-1)\over t_0^2},\qquad \alpha = \pm\sqrt{s+1\over 2s+1}
\end{equation}
The cosmological scale factor exponent runs over the disjoint range $\{s\ge 0\}\cup \{s\le -1\}$, with the parameter 
$\epsilon$ in (\ref{fonarev}) equal to either $+1$ or $-1$ respectively in the two parts of the range.  With $r_0=0$, the Fonarev spacetimes reduce to the scalar FRW cosmologies described in the Appendix.  For $r_0\neq 0$ they are clearly asymptotic to these cosmologies for $r\gg r_0$, and it again seems reasonable to consider them to represent JNW-type inhomogeneities embedded in these cosmological backgrounds.
If we consider fixing the parameters $\lambda$, $V_0$ that determine the scalar potential in the action (\ref{action}), we see that the cosmological scale factor exponent $s$ and time scale $t_0$ are then determined, as are the possible values of the JNW parameter $\alpha$.  The HMN spacetimes (\ref{hmn}) correspond to the special case $s=1/2$ for which the scalar potential vanishes (in addition to $s=0$ which corresponds to the original, static JNW spacetimes).

\subsection{Generalized dimensional reduction}

In this section we review the generalized dimensional reduction scheme introduced in reference \cite{Gouteraux:2011qh}.  One starts with the action for Einstein gravity with a possible cosmological constant in $4+n$ dimensions 
\begin{equation}\label{higherd}
S_{4+n} = \int d^{4+n}x\sqrt{-g_{(4+n)}}\left( R_{(4+n)}-2\Lambda\right)
\end{equation}
One now assumes that $ds_{4+n}^2$ solves the corresponding equations of motion and has the form
\begin{equation}\label{reduce}
ds_{4+n}^2 = e^{-n\sigma\phi / 2}ds_4^2 + e^{\sigma\phi}ds_n^2
\end{equation}
where $ds_4^2$ is the metric on a $4$ dimensional space ${\cal M}_4$, $ds_n^2$ is the metric on an $n$-dimensional Einstein space ${\cal M}_n$ with constant scalar curvature $R_{(n)}$, $\phi$ is a scalar field that depends only on the coordinates on 
${\cal M}_4$, and
\begin{equation}
 \sigma= {2\over\sqrt{n(n+2)}}
\end{equation}
The $4D$ metric and scalar field will then solve the equations of motion that follow from the $4D$ action
\begin{equation}\label{genreduce}
S= \int d^4x\sqrt{-g}\left (R - {1\over 2}(\nabla\phi)^2 - V_0^{(1)} e^{-\lambda_1\phi}
-V_0^{(2)} e^{-\lambda_2\phi}\right)
\end{equation}
where the parameters of the scalar potential terms are given by $V_0^{(1)}=2\Lambda$, $V_0^{(2)}=-R_{(n)}$, $\lambda_1=n\sigma/2 $, and
 $\lambda_2= 2/ (n\sigma ) $.  The ``generalized" aspect of this dimensional reduction scheme is that having established this connection for positive integer values of $n$, one considers it to hold for continuous values of $n$ as well.  This becomes useful if one starts with a family of higher dimensional solutions that is known for all dimensions through an analytic expression in $n$.  We will assume that the number of generalized extra dimensions $n\ge 0$, although some of the formulas also turn out to work with $n<0$.

 In the generic case, one would have both coefficients $V_0^{(1)}$ and $V_0^{(2)}$ in the scalar potentials nonzero.  However, we will follow \cite{Gouteraux:2011qh} and focus on the cases when, in the higher dimension, only one of either $\Lambda$ or $R_{(n)}$ is non-vanishing, which will imply that either one, or the other of the scalar potential terms vanishes.  This will lead to two different schemes for obtaining the cosmological JNW spacetimes;  one in which the higher dimensional $\Lambda\neq 0$, but the $n$-dimensional compactifying space has vanishing scalar curvature; and a second scheme in which $\Lambda=0$, but the compactifying space is curved.  In each of these two schemes, the four dimensional action then reduces to the form (\ref{action}), {\it i.e.} with only a single exponential potential interaction term for the scalar field.

\subsection{Scalar FRW cosmologies via generalized dimensional reduction}\label{general}

The cosmological JNW spacetimes (\ref{fonarev}) asymptote at large radius to the FRW cosmological solutions of the theory (\ref{action}), which are explored in the Appendix.  In this section we will show how these homogeneous cosmological solutions may be obtained from generalized dimensional reduction of solutions to the higher dimensional theory (\ref{higherd}).  As noted above, this may be done in two ways, which cover complementary portions of the parameter space of FRW solutions.

In what we will call \textsl{Scheme A},  the cosmological constant  in the higher dimensional action (\ref{higherd}) is taken to be positive, $\Lambda\ge 0$, while the scalar curvature of the compactification space is taken to vanish.  Using the results from the last subsection, the compactified system is then described by the $4$ dimensional action (\ref{action}) with the parameters of the scalar potential given by $V_0=2\Lambda$ and $\lambda= \sqrt{n/(n+2)}$.
As a solution to the higher dimensional action (\ref{higherd}), we take $4+n$ dimensional deSitter spacetime, which can be written in terms of a conformal time coordinate as
\begin{equation}\label{bigmetric}
ds_{4+n}^2 = (t / t_0 )^{-2} \left(-dt^2 + \delta_{ij}dx^idx^j+\delta_{\alpha\beta}dy^\alpha dy^\beta \right)
\end{equation}
where $i,j=1,2,3$ and $\alpha,\beta = 1,\dots,n$.  The $n$ dimensional Einstein space ${\cal M}_n$ is simply the flat space spanned by the coordinates $y^\alpha$. Dimensional reduction using (\ref{reduce}) then yields the $4$ dimensional fields
\begin{equation}\label{scalar1}
ds_4^2 = (t /t_0 )^{-(n+2)}(-dt^2 +\delta_{ij}dx^idx^j),\qquad e^\phi=(t/t_0)^{-\sqrt{n(n+2)}}
\end{equation}
where $t_0^2 = (n+2)(n+3)/V_0$.
This coincides with the scalar cosmologies (\ref{FRW}) with scale factor exponent $s=-(n+2)/2$.  Given that the number of generalized extra dimensions is assumed to be positive, $n\ge 0$, the scalar FRW cosmologies produced by \textsl{Scheme A} therefore cover the range\footnote{Note that the 
scalar field in (\ref{scalar1}) will also be real if $n\le -2$, which corresponds  to scale factor exponents in the range $s\ge 0$.  However, the sign of the exponent for the scalar field in (\ref{scalar1}) is incorrect for $n<0$, indicating that some type of additional analytic continuation argument may be required to make sense of generalized dimensional reduction with $n<0$.} $s\le -1$.  

FRW cosmological solutions of (\ref{action}) can also be obtained by means of a \textsl{Scheme B}.  In order to distinguish between the two schemes, we now start  in $4+n^\prime$ dimensions, taking
$\Lambda=0$ in the higher dimensional action (\ref{higherd}) and compactifying on an Einstein space ${\cal M}_{\np}$ with constant negative scalar curvature, $R_{({\np})}<0$.  In this case, the parameters of the $4$ dimensional scalar potential in (\ref{action}) are given by $\lambda =\sqrt{(\np+2)/\np}$ and $V_0=-R_{(\np)}$.
We now start with Minkowski spacetime in $4+\np$ dimensions
\begin{equation}
ds_{4+\np}^2 = -d\tau^2 + \delta_{ij}dx^idx^j+\delta_{\alpha\beta}dy^\alpha dy^\beta
\end{equation}
and transform to Milne coordinates in the $\np+1$ dimensional Minkowski subspace spanned by the $(\tau,y^\alpha)$ coordinates to obtain\footnote{See {\it e.g.} reference \cite{Russo:2003ky} for the explicit form of the transformation of Minkowski spacetime into Milne coordinates.}
\begin{equation}\label{flatmilne}
ds_{4+\np}^2 = -dt^2 + \delta_{ij}dx^idx^j+(t/t_0)^2dH_{\np}^2
\end{equation}
where $dH_{\np}^2$ is a metric on a constant curvature $n$ dimensional hyperbolic space\footnote{This can be made compact by identifying coordinates via a discrete symmetry.  However, this is not important for the present construction.} ${\cal H}_{\np}$, which in particular space has constant scalar curvature $R_{({\np})}\le 0$.  Dimensionally reducing the metric (\ref{flatmilne}) using (\ref{reduce}) gives the $4D$ fields
\begin{equation}
ds_4^2 = (t/t_0)^{\np}(-dt^2 +\delta_{ij}dx^idx^j),\qquad e^\phi=(t/t_0)^{\sqrt{{\np}({\np}+2)}}
\end{equation}
where $t_0^2= {\np}({\np}-1)/V_0$.
This reproduces scalar cosmologies (\ref{FRW}) with scale factor exponent $s=\np/ 2$, and we see that  \textsl{Scheme B} with a positive number of generalized extra dimensions, $\np\ge 0$, produces scale factor exponents in the range $s\ge 0$. 
This precisely complements the range of exponents covered by \textsl{Scheme A}. 
The $5D$ Kaluza-Klein warm-up presented in Section (\ref{hmnkk}) corresponds to a limiting case of \textsl{Scheme B} with only a single extra dimension, {\it i.e.} $\np=1$, so that the space $H_{\np}$ is simply the real line, which has $R_{({\np})}=0$.  The expression for $t_0^2$ given above can remain finite in this limit because both the numerator and denominator are approaching zero.

\subsection{JNW branes}

We now want to use these two generalized dimensional reduction schemes (\textsl{Scheme A} in which we take $\Lambda>0$  in the higher dimensional space, and \textsl{Scheme B} in which we compactify on a hyperbolic space with $R_{(n)}< 0$)  to construct the $4D$ cosmological JNW spacetimes.  In order to do this we need to find suitable solutions to the higher dimensional action (\ref{higherd}) to work from.  For both schemes, these will be based on higher dimensional versions of the JNW spacetimes (\ref{jnw}) which we will call JNW branes.

Recall that in section (\ref{hmnkk}) we lifted the $4D$ JNW spacetimes, which are solutions to Einstein gravity coupled to a massless scalar field, to $5D$ solutions to the vacuum Einstein equations.  The resulting $5D$ spacetimes (\ref{JNW_uplift}) can be thought of as JNW strings, extended in a translationally invariant way along a single additional spatial direction.  More general JNW branes will be translationally invariant along additional `world-volume' directions as well.  Based on the JNW string solutions, we adopt the following ansatz for $N+3$ dimensional JNW branes, which is translationally invariant in the $N$ directions tangent to the brane and spherically symmetric in the $3$ dimensions orthogonal to the brane,
\begin{equation}\label{jnwbrane}
ds_{N+3}^2 = f^p\left(dr^2 + fr^2d\Omega^2\right) +\sum_{i=1}^N\epsilon_i f^{q_i} (dx^i)^2,\qquad f=1-{r_0\over r}
\end{equation}
Here  $\epsilon_i=\pm 1$, allowing for both Euclidean and Lorentzian branes, both of which will be useful below.   
One finds that the JNW brane ansatz (\ref{jnwbrane}) satisfies the vacuum Einstein equations  if  the exponents $(p,q_i)$ obey the Kasner-like constraints
\begin{equation}\label{jnwconstraints}
p +\sum_{i=1}^N q_i = 0,\qquad p^2+\sum_{i=1}^N q_i^2 = 2
\end{equation}
 We will be particularly interested in the case where all the $q_i$ are equal to a common value $q$, in which case the constraints (\ref{jnwconstraints}) reduce to two equation in two unknowns.  
Then the solutions to (\ref{jnwconstraints}) are  
\begin{equation}\label{pandq}
p =\mp\sqrt{{2N\over N+1}},\qquad q = \pm\sqrt{{2\over N(N+1)}}
\end{equation}
Static JNW branes (\ref{jnwbrane}) with exponents (\ref{pandq}) will serve as starting points for our constructions of  the cosmological JNW spacetimes using generalized dimensional reduction.  It is important, in this respect, that the exponents (\ref{pandq}) are given by analytic expressions in the number of dimensions, so that  $N$ can be treated as a continuous variable.

For orientation purposes, it is useful to look at some simple examples of JNW branes.  With $N=1$, there are two solutions to the constraints (\ref{jnwconstraints}) given by
\begin{equation}
(p,q_1) = (-1,+1),\qquad (p,q_1)=(+1,-1)
\end{equation}
The first of these gives Schwarzschild spacetime in its usual form, either Euclidean or Lorentzian depending on the choice of $\epsilon_1$.  The second choice can also be recognized as Schwarzschild by transforming (\ref{jnwbrane}) to the new radial coordinate $r^\prime = r_0-r$.

We have already encountered (Lorentzian) $N=2$ JNW branes  when we lifted the JNW spacetimes to $5D$ vacuum solutions in Section (\ref{hmnkk}) resulting in (\ref{JNW_uplift}).
Rewriting these spacetimes in the form of the JNW brane ansatz (\ref{jnwbrane}), we have $\epsilon_1=-1$ and $\epsilon_2=+1$, and exponents
\begin{equation}\label{5Dexponents}
p= \mp\sqrt{{1-\alpha^2\over 3}} - \alpha,\qquad
q_1= \mp\sqrt{{1-\alpha^2\over 3}} + \alpha,\qquad
q_2= \pm 2\sqrt{{1-\alpha^2\over 3}}
\end{equation}
and one can check that these indeed satisfy the Kasner-like constraints (\ref{jnwconstraints}).  The requirement imposed in Section (\ref{hmnkk}) of being able to transform to Milne coordinates is the same as that imposed in (\ref{pandq}).
 
 \subsection{Cosmological JNW spacetimes - Scheme A}
 
We now proceed to build the cosmological JNW spacetimes (\ref{fonarev}) via \textsl{Scheme A} for generalized dimensional reduction and making use of the vacuum JNW brane solutions.  \textsl{Scheme A} involves starting with a solution to the higher dimensional action (\ref{higherd}) with $\Lambda>0$ and in order to generate the scalar FRW cosmologies via this scheme, we started with $n+4$ dimensional deSitter spacetime.  We can now add in a JNW-type inhomogeneity to this deSitter background in the following way.  Recall that if we have a $(3+n)$-dimensional Euclidean solution to the vacuum Einstein equations with line element $ds_{3+n}^2$ then the $4+n$ dimensional spacetime 
\begin{equation}\label{oneup}
ds_{4+n}^2 = (t/t_0)^{-2 }(-dt^2 +ds_{3+n}^2)
\end{equation}
will solve the Einstein equations\footnote{A corresponding construction is commonly used in the AdS context, see {\it e.g.} reference \cite{Chamblin:1999by}.} with cosmological constant $\Lambda = (n+2)(n+3)/2 t_0^2$.  
If the $(3+n)$-dimensional metric is taken to be flat, then the resulting $(4+n)$-dimensional spacetime (\ref{oneup}) is deSitter spacetime, giving back the starting point for the construction of the scalar FRW cosmologies via \textsl{Scheme A}.
We can add a JNW-type inhomogeneity by taking $(3+n)$ dimensional metric in (\ref{oneup}) to instead be a Euclidean JNW brane (\ref{jnwbrane}) 
with all the $q_i$ equal, so that $p$ and $q$ are given as in (\ref{pandq}).  The $(4+n)$-dimensional metric (\ref{oneup}) is then asymptotic to deSitter spacetime at large $r$  and is given by
\begin{align}\label{wlambda}
ds_{4+n}^2 =  & (t/t_0)^{-2}\left (-dt^2 + f^{ -2\alpha } (dr^2 +r^2 f d\Omega^2) + f^{ 2\alpha/n}\sum_{i=1}^n (dx^i)^2\right) \\
f=& 1-{r_0\over r} ,\qquad \alpha =  \pm \sqrt{{n\over 2( n+1)}} \nonumber
\end{align}
Dimensional reduction using equation (\ref{reduce})  now yields the $4D$ spacetime
\begin{equation}
ds_4^2 = (t/t_0)^{-(n+2)}\left ( -f^{\alpha} dt^2 + f^{- \alpha}(dr^2 +r^2 f d\Omega^2)\right),\qquad 
e^\phi = (t/t_0)^{\sqrt{n(n+2)}}f^{\pm\sqrt{1-\alpha^2}}
\end{equation}
with $\alpha$ given as above taking values in the range $\alpha^2\le 1/2$ for positive values of $n$.  This reproduces the cosmological JNW solutions (\ref{fonarev}) found by Fonarev \cite{Fonarev:1994xq} for this range of JNW parameter.

One particularly interesting case of the above construction is when $\alpha=1/2$ corresponding to $n=1$.  The higher dimensional space is then $5$ dimensional and the metric (\ref{wlambda}) is given by
\begin{equation}\label{soliton}
ds_{5}^2 =   (t/t_0)^{-2}\left (-dt^2 + f dx^2 + f^{ -1 } dr^2 +r^2 d\Omega^2 \right)
\end{equation}
The spatial part of the metric, if we ignore the overall time dependent conformal factor, is simply the Euclidean Schwarzschild geometry, which closes off smoothly at $r=r_0$ if the period of the coordinate $x$ is chosen appropriately and otherwise has a conical singularity there.  Including the time direction, but still ignoring the conformal factor, we see the static Kaluza-Klein bubble, a spacetime which has standard Kaluza-Klein boundary conditions at infinity, but with the extra dimension shrinking down to zero size in the interior, so that the entire space cuts off at $r=r_0$.  The static Kaluza-Klein bubble dimensionally reduces via (\ref{kk_diagonal}) to the JNW spacetime with $\alpha=1/2$.
Including the time dependence, we now have a deSitter version of the Kaluza-Klein bubble\footnote{The spacetime can also be recognized as an analytic continuation of the AdS string \cite{Chamblin:1999by}.}.  The feature of interest for us is simply that the curvature is well behaved at $r=r_0$ for both the static and deSitter versions of the Kaluza-Klein bubble, giving nonsingular higher dimensional origin for the static and cosmological JNW solutions with $\alpha = 1/2$.

 \subsection{Cosmological JNW spacetimes - Scheme B}

The \textsl{Scheme B} construction of the cosmological JNW spacetimes
 makes use of Milne coordinates for the Minkowski metric in order to introduce a compactification space with constant negative curvature.  In order to include a JNW-type inhomogeneity, we start in $4+\np$ dimensions with a JNW-brane solution that allows us to make a similar transformation to Milne coordinates in an appropriate subspace.  This is done by taking $N=\np+1$ in the JNW brane solution (\ref{jnwbrane}), incorporating a  time directionby taking $\epsilon_{\np+1}=-1$ and relabeling $x^{\np+1}=\tau$, and again taking all the exponents $q_i$ to be the same and therefore given by (\ref{pandq}).   The starting metric is then given by
\begin{equation}
ds_{4+\np}^2 = f^p(dr^2 + fr^2d\Omega^2) + f^{-p/(\np+1)}(-d\tau^2 +\sum_{i=1}^{\np}(dx^i)^2),\qquad 
p= \sqrt{ {2(\np+1)\over  \np+2 } }
\end{equation}
and has $\np+1$ dimensional boost invariance.
We can therefore transform to Milne coordinates in the subspace spanned by the coordinates $(\tau,x^i)$ resulting in the metric
\begin{equation}
ds_{4+\np}^2 = f^p(dr^2 + fr^2d\Omega^2) + f^{-p/(\np+1)}(-dt^2  +(t/t_0)^2 dH_{\np}^2)
\end{equation}
where again $dH_{\np}^2$ is the line element for a hyperbolic space with constant  negative curvature, such that $R_{(\np)}=-\np(\np-1)/ t_0^2$.  Dimensional reduction via (\ref{reduce}) then yields the 4 dimensional fields
\begin{equation}
ds_4^2 = (t/t_0)^{\np}\left\{ -f^\alpha dt^2 + f^{-\alpha} (dr^2 +r^2 f d\Omega^2)\right\} ,\qquad
e^\phi = (t/t_0)^{\sqrt{\np(\np+2)}}f^{\pm\sqrt{1-\alpha^2}} 
\end{equation}
where now the JNW parameter is related to the number of generalized extra dimensions $\np$ according to
\begin{equation}
\alpha   = \pm  \sqrt{ { \np+2 \over 2(\np+1)}}
\end{equation}
These are again cosmological JNW spacetimes (\ref{fonarev}) and we see that with $\np\ge 0$, the results of \textsl{Scheme B} cover the range of JNW parameter $1/2\le\alpha^2\le 1$.  This complements the range of solutions produced by \textsl{Scheme A} precisely, so that between the two the entire range of cosmological JNW solutions is reproduced via generalized dimensional reduction.

\section{Discussion}\label{discussion}
We have shown that the $4D$ cosmological JNW spacetimes (\ref{fonarev}) can be obtained, starting from the higher dimensional JNW brane solutions (\ref{jnwbrane}), via two complementary generalized dimensional reduction schemes, which together cover the entire parameter space of solutions. 
Knowledge of exact dynamical, cosmological black hole solutions would be useful for understanding the formation, evolution and possible roles played by of black holes in the early universe. The cosmological JNW spacetimes \cite{Husain:1994uj,Fonarev:1994xq} do not themselves succeed in providing a useful model \cite{Maeda:2007bu} in this regard. 
However, the higher dimensional construction of these spacetimes may nonetheless provide a valuable clue for how to build actual regular, dynamical  cosmological black hole solutions. 
The reason for this optimism is that 
the JNW spacetimes represent a peculiar neutral limit of charged dilaton black holes \cite{Gibbons:1987ps,Garfinkle:1990qj}, the magnetically charged versions of which are given by %
\begin{align}\label{dilaton}
ds^2 & = -Fdt^2 +F^{-1}dr^2 +R^2d\Omega^2,\qquad A_\varphi = q\cos\theta,\qquad e^{-a\phi}=(1-{r_-\over r})^{{2a^2\over 1+a^2}}\\
F&= (1-{r_+\over r})(1-{r_-\over r})^{{1-a^2\over 1+a^2}},\qquad R^2=r^2(1-{r_-\over r})^{{2a^2\over 1+a^2}}\nonumber
\end{align}
and have mass and charge 
\begin{equation}
M= {r_+\over 2}+({1-a^2\over 1+a^2}){r_-\over 2},\qquad Q=({r_+r_-\over 1+a^2})^{1\over 2}
\end{equation}
Strictly speaking these are black hole solutions only if one takes $r_+>r_-$, because the curvature is singular at $r_-$.  The standard neutral limit is achieved by taking $r_-=0$, which gives Schwarzschild for all values of the dilaton coupling $a$.  However, taking $r_+=0$ gives a second neutral limit and this yields the JNW spacetime (\ref{jnw}), with the JNW parameter related to the dilaton coupling according to $\alpha=(1-a^2 ) /( 1+a^2)$.  This again reduces to Schwarzchild when $a=0$.
This is interesting because  solutions for BPS dilaton black holes with $r_+=r_-$ embedded in the scalar cosmologies (\ref{FRW}) are also known \cite{Maki:1992tq}.  Hence the cosmological JNW metrics
can be thought of as a second data point that can be applied to the puzzle of constructing cosmological versions of the entire family
 of dilaton black hole solutions (\ref{dilaton}).

Some further remarks are as follows.  We have noted that the family of static JNW spacetimes (\ref{jnw}) are singular at $r_0$, with the exception of the Schwarzschild black hole, $\alpha =1$, which has vanishing scalar field. This includes the case
$\alpha =1/2$, which we have found is the reduction of a smooth $5D$ Kaluza-Klein bubble spacetime. This is true for the cosmological JNW solutions with $\alpha=1/2$ as well.
The dimensional reduction, therefore,  does not
accurately capture the smooth $5D$ bubble geometry. Possibly there is a more sophisticated dimensional reduction which gives a regular $4D$
spacetime that is derived from a modified $4D$ action. The issue in this case is that the $5D$ coordinates are not well defined at $r_0$, which describes the origin of an $\mathcal{R}^2$ plane in 
in polar coordinates. The dimensional reduction prescription (\ref{reduce}) is then not valid
at the location of the bubble. A similar issue arises when computing the Euclidean action of a black hole in the Hamiltonian framework, where   a
proper treatment yields a contribution to the action proportional to the area of the Euclidean horizon \cite{Hawking:1995fd}. In future work, it would be interesting
to study dimensional reduction including the possible contributions from such an inner boundary term. 

Lastly, we turn briefly to the issue of regularity of the spacetimes by looking at trapped surfaces, cosmological horizons, and
the higher dimensional constructions of the cosmological JNW spacetimes that we have explored here. 
Trapped surfaces for outward (inward) propagating null rays  are compact co-dimension two surfaces on which the expansion of outward (inward)
propagating null rays, $\theta _{\pm} $, vanishes.
For example, in a Schwarzchild-deSitter spacetime,  $\theta_+ =0$ on the black hole horizon
 and $\theta _- =0$ on the future cosmological horizon. On the other hand, on the white hole horizon $\theta _- =0$ and
on the past cosmological horizon $\theta _+ =0$.
 A thorough analysis of the trapped surfaces of the 4$D$ cosmological JNW spacetimes was done in \cite{Maeda:2007bu}. 
 The details vary depending on the value of $\alpha$, but generically contracting (expanding) cosmologies have black-hole type (white-hole type) trapped surfaces.
  Accelerating (decelerating)  cosmologies have future (past) cosmological horizons. It was found, however, that black hole trapped surfaces
 exist only for a certain portion of the evolution, so that the singularity at $r_0$ is eventually visible. The conclusion was that the
  cosmological JNW spacetimes are not satisfactory descriptions of cosmological black holes.
  
  It is natural to ask if the situation is improved in the higher dimensional spacetimes. Generically, the answer is no.  Consider spherically symmetric 
  surfaces with unit outward normal $r^a$, in a constant time slice. The trapped surface equation 
 can be written as  $\theta _\pm = \pm D_a r^a + K^2 -K_{ab}K^{ab} =   0$ , where $K_{ab}$ is the extrinsic  curvature of the slice in the spacetime. Though the individual pieces
 in this equation are not the same,  the equations reduce to 
 the same requirement in the higher dimensional spacetime as in its 4$D$ reduction. Likewise, as long as null rays have no components 
 in the direction of the internal coordinates, the null paths are the same in the higher and lower dimensional spacetimes since those chunks of the
 metric are conformally related. So the analysis of null paths in the  higher dimensional spacetimes is the same as in 4$D$, though the geometrical 
 properties of the boundaries can differ.

\appendix

\vskip 0.2in
{\huge{\bf Appendix}}

\section{Scalar FRW cosmologies}\label{appendix}

In this paper we are interested in JNW-type inhomogeneities embedded in FRW cosmological backgrounds that are solutions to Einstein gravity coupled to a scalar field with an exponential potential (\ref{action}).  
Written in terms of a conformal time coordinate, these FRW cosmological solutions \cite{Halliwell:1986ja,Barrow:1987ia,Yokoyama:1987an}  are given by  
\begin{equation}\label{FRW}
ds^2 =   (t/t_0)^{2s}(-dt^2+\delta_{ij}dx^idx^j)\ ,\qquad e^\phi = (t/t_0)^{2\epsilon\, \sqrt{s(s+1)}}
\end{equation}
where $\epsilon =\pm 1$ and the parameters of the scalar field potential are related to the scale factor exponent $s$ and the time scale $t_0$ according to
\begin{equation}
\lambda = \sqrt{{s+1\over s}},\qquad V_0={2s(2s-1)\over t_0^2}
\end{equation}
In order that the exponent $\lambda$ be real, the scale factor exponent $s$  is restricted to run over the disjoint ranges  $s\le -1$ and $s\ge 0$.  One also finds that the sign of the scalar field exponent in (\ref{FRW}) is given by $\epsilon=-1$ for $s\le -1$ and $\epsilon=+1$ for $s\ge 0$.   For $s=1/2$, we see that the scalar potential vanishes, corresponding to a massless scalar field.  This gives the cosmological background for the HMN solutions (\ref{hmn}).  The exponent $\lambda$ in the scalar potential vanishes for $s=-1$, which is deSitter spacetime.

These same solutions written in terms of a proper time coordinate have the form
\begin{equation}
ds^2 =   -dT^2+ (T/ T_0 )^{2p}\delta_{ij}dx^idx^j \ ,\qquad e^\phi =(T/T_0)^{2\sqrt{p}}
\end{equation}
where the parameters of the scalar potential are now given in terms of the scale factor exponent by 
\begin{equation}
\lambda={1\over\sqrt{p}},\qquad V_0T_0^2=2p(3p-1).
\end{equation}
The proper and conformal time coordinates are related through the transformation
$T/T_0 = (t/ t_0)^{s+1}$ with $T_0= t_0 / s+1 $ and the scale factor exponents are related according to
\begin{equation}\label{exponents}
s = {p\over 1-p}, \qquad  p={s\over s+1}
\end{equation}
We see that $s\ge 0$ corresponds to $0\le p<1$, while $s\le -1$ corresponds to $p>1$.
The expansion rate of the universe is accelerating if $p>1$, or in terms of the conformal time coordinate  $s\le -1$ range. 
Note that the expanding phase of (\ref{FRW}) for $s\le -1$ corresponds to the conformal time coordinate running between $-\infty$ and $0$.  
 
The case of deSitter spacetime, with $s=-1$ which corresponds to the limit $p\rightarrow\infty$, requires special handling. 
In terms of the proper time coordinate we have
\begin{equation}
ds^2 = -dT^2 +e^{2T/T_0}\delta_{ij}dx^idx^j
\end{equation}
where $ e^{T/T_0 } = t_0  / t $ and $T_0=-t_0$, to that $t_0<0$ corresponds to the exponentially expanding region of deSitter.

\end{document}